\documentstyle[11pt,newpasp,twoside,epsf]{article}
\begin{document}

\title{New Pulsars from Arecibo Drift Scan Search}

\author{ A. Somer}
\affil{ Astronomy Department \& Radio Astronomy Laboratory,
University of California, Berkeley, CA
}

\begin{abstract}

We report the discovery of 3 pulsars, PSR J0030+0451, PSR J0711+09, and
PSR J1313+09 that were found in a three dimensional  
(DM, period, position) search at 430 MHz
using the 305m Arecibo telescope.  
PSR J0030+0451 is a nearby 4.8-millisecond solitary pulsar. 
Spin and astrometric parameters are presented for the three new pulsars.
We have measured significant polarization in
the millisecond pulsar, PSR J0030+0451, over more than 50\% of the period
and use these data and also a morphological decomposition of the profile
to briefly discuss magnetospheric models.
\end{abstract}

\section {Introduction}

From 1994 to 1995, during the time of the Gregorian Dome upgrade at the
Arecibo Observatory, while the telescope had limited steerability due to
construction, a joint effort by several collaborations was aimed at
surveying the sky visible from Arecibo 
(declination range roughly -1$\deg$ to 39$\deg$) 
in search of new pulsars. 
A total of 44 pulsars including 5 MSPs have been discovered so far
by the other institutions (Foster et al. 1995, Camilo et al. 1996, Ray et al. 1996).
Our search has an estimated sensitivity of 0.9 mJy at 7$\sigma$
for long period, low DM, low zenith angle, high galactic latitude pulsars
which is comparable to previous searches.
In December 1997 we confirmed the presence of 3 new pulsars: J0030+0451 (4.8ms), J0711+09
(2.4s), and J1313+09 (0.85s) at the Arecibo Observatory using the
Arecibo Berkeley Pulsar Processor and the Penn State Pulsar Machine 
(Backer et al. 1997, Foster et al. 1995).
Follow up observations of the 3 new pulsars
were conducted over a period of nearly 2 years from 
December 1997 through September 1999 at the
Arecibo Observatory.

\vskip 2 in

\section{Analysis}

The data were cross correlated with a template, 
and the resulting TOAs analyzed using the TEMPO program (Taylor \& Weisberg 1989).
Table 1 provides spin, astrometric and other parameters for each of the 
3 new pulsars.  

\begin{table}
\caption[Pulsar Parameters] {Observed pulsar parameters.} 
\begin{tabular}{||l|c|c|c||} \tableline
     &				J0030+0451   &   J0711+09   & J1313+09 \\ \tableline
Right Ascension (2000) &		00 30 27.4339(6)	  &  07 11 36.74(2.00) & 13 13 26.5(2.0) \\ \tableline					
Declination (2000) &		04 51 39.65(2)  &  09 31 40(30) & 09 33 31(30) \\ \tableline
Galactic longitude & 113.1$\deg$ & 206.7$\deg$  & 320.4$\deg$ \\ \tableline
Galactic latitude & -57.6$\deg$    &  8.8$\deg$ & 71.7$\deg$ \\ \tableline
Period(s) & 	0.00486545320737(1) & 1.21409045(4) & 0.84893276(7) \\ \tableline
$\dot{P}$ $(10^{15})$ s s$^{-1}$&   1.0(2)$\times 10^{-5} $ &  0.246(300)    &  1.947(2000) \\ \tableline
Epoch (MJD) &				50984.4 &  50699.5  &         50984.5 \\ \tableline
Dispersion Measure & & & \\(pc cm$^{-3}$) &            4.3326(1) & 46.15(200)      &   12.0(1) \\ \tableline
Timing Data Span (MJD) &		50789-51277 & 50789-51104  & 50788-51080 \\ \tableline
RMS timing residual $(\mu$s)\tablenotemark{a} & 13       & 4500           &  1400 \\ \tableline
Flux $S_{400}$ (mJy) &		7.9(1)		& 2.4(1)  &       3.5(1)  \\ \tableline
Flux $S_{1400}$ (mJy) &		0.6(1)    & $>$ 0.04(1) &  $>$ 0.16(1)  \\ \tableline
DM distance (pc)\tablenotemark{b} &		230 & 2450  & 780 \\ \tableline
$L_{400}$ mJy kpc$^{2}$ & 0.4  & 14 & 2.1\\ \tableline
Spectral index & $2.2 \pm .2$ & $< 3.5 $ & $< 2.6$ \\ \tableline
Characteristic age (y) &  $ 8 \times 10^9 $ &  $ 7 \times 10^7 $ & $7\times 10^6 $ \\ \tableline
Magnetic field (G)\tablenotemark{c} & $ 2.2 \times 10^8 $ & $5.5 \times 10^{11} $ & $ 1.3 \times 10^{12} $\\ \tableline 
Proper motion upper & & &\\
limit (arcsec/yr) & .060 & (none) &  (none)   \\ \tableline
\end{tabular} 
\tablenotetext{a}{RMS from 3 minute averages}
\tablenotetext{b}{Model from Taylor \& Cordes, 1993} 
\tablenotetext{c}{$ B_o = 3.2 \times 10^{19} \sqrt{P\dot{P}} $}
\end{table}

\section {Millisecond Pulsar : PSR J0030+0451}

A pulsar displaying an interpulse like PSR J0030+0451 has two distinct
possible geometries:  nearly aligned, with the interpulse resulting from
the second crossing of a wide-angle hollow ``cone" of radiation, or 
orthogonal, with the two emission regions coming from opposite poles.
We performed a dual frequency morphological study of the intensity
profile to look for clues to the geometry of the system.  Figure 1
depicts the Gaussian decomposition of the profile at each frequency,
with the corresponding Gaussian parameters shown in Table 2 
(Kramer et al. 1994, Kramer 1994).  
Conal emission is suggested by the narrowness of the 1400 MHz model
compared to the 430 MHz data (panel b overlays the two).
In addition, the amplitudes of the Gaussian components
suggest that possibly all but component 2 are steep spectral index peaks, 
which may indicate that
component 2 is a core component, and the rest are cone components.

The polarization data do not settle the question of whether the
geometry is orthogonal or
nearly aligned.  Figure 2 shows an example of each.  

\begin{figure}
\plotone{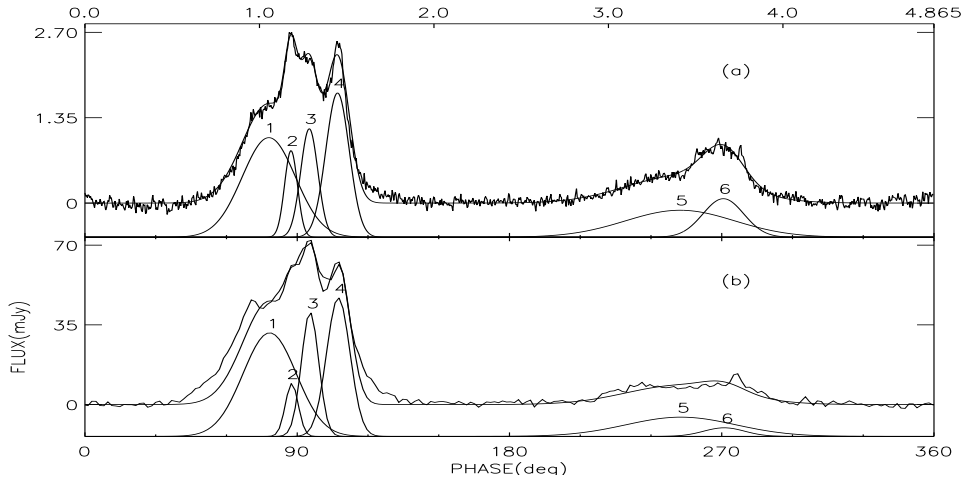}
\caption{Decomposition of PSR J0030+0454 intensity profile into 6 components.  
(a) 1400 MHz (b) 430 MHz.  The best-fit Gaussian components at 1400 MHz
were only allowed to vary in amplitude in order to fit the 430 MHz profile.
(cf. Lorimer, ``A Week in Review'', these proceedings)}
\end{figure}

\begin{table}
\caption[Table 2]{Fitted Gaussian parameters for the 6 components in Figure
2. The amplitude of component 2 was arbitrarily set to 1.0 for both frequencies.  All other amplitudes are relative to component 2.
}
\begin{tabular}{|l|r|r|r|r|r|r|r|}
\tableline
Peak \# & 1 &    2 &     3 &     4 &     5 &     6 \\
\tableline
Center (deg) &78.74 & 88.13 & 95.88 & 107.85 &  253.106 & 271.45 \\
Width (deg) & 15.96 & 3.854 & 5.183 & 7.012 &   32.77 & 12.23 \\
Amplitude (430 MHz) & 1.95 & 1.00 & 2.34 & 2.61 &  0.771 &  1.26  \\    
Amplitude (1400 MHz) & 1.15  & 1.00 & 1.25 & 1.67 & 0.473 &  0.442 \\
\tableline
\end{tabular}
\end{table}

\begin{figure}
\plotone{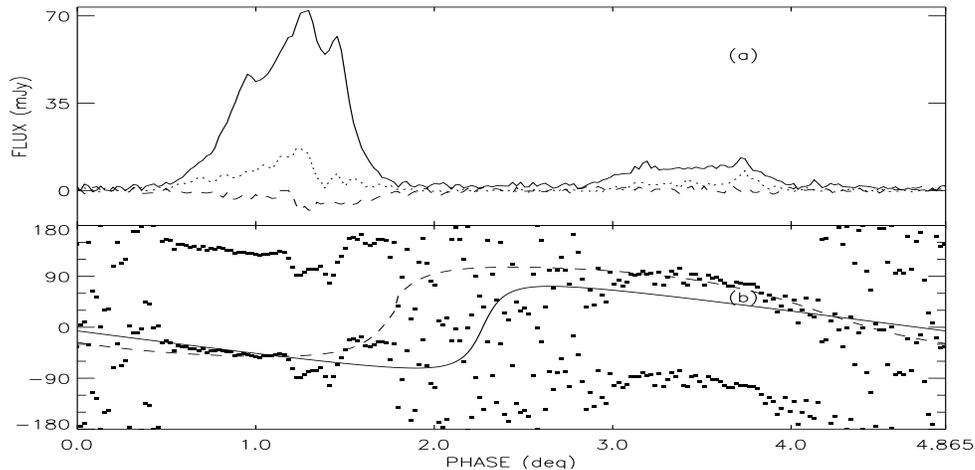}
\caption{PSR J0030+0454 at 433 MHz.  (a) 
The solid line shows intensity vs. phase.  The dotted line shows
linear polarization and the dashed line shows circular polarization.   
(b) Two {\it RVM}s are superimposed on the PPA data.
The solid line demonstrates a possible
nearly aligned rotator,
plotted with
$ \alpha=8 \deg $ and $ \beta =1 \deg $.  
The dashed line demonstrates a possible 
orthogonal rotator with values
$ \alpha=62 \deg $ and $ \beta =10 \deg $. }  
\end{figure}

Of the 36 known MSPs in the disk of the galaxy, 
PSR J0030+0451 is the 9th solitary MSP, i.e., it is not
in a binary system.  These objects present a unique problem in the standard
evolutionary
scenario of MSPs.  If the neutron star is spun up via accretion
by a mass transfer from a companion star, the companion must somehow be 
obliterated.  

\section {Acknowledgements}
I'd like to thank the LOC for their financial support 
and for a wonderful conference!
Many thanks to Michael Kramer for timing observations from
the MPIfR 100-m telescope and for the use of his Gaussian fitting program,
Alex Wolszczan for use of the Penn State Pulsar Processor, 
and
Kiriaki Xilouris and Duncan Lorimer for extensive assistance in observations.


\begin{references}

\reference Backer, D.~C., Dexter, M.~R., Zepka, A., Ng, D., Werthimer, D.~J., Ray, P.~S.,
  \& Foster, R.~S. 1997, PASP, 109, 61

\reference Camilo, F., Nice, D.~J., Shrauner, J.~A., \& Taylor, J.~H. 1996, ApJ, 469, 819

\reference Foster, R.~S., Cadwell, B.~J., Wolszczan, A., \& Anderson, S.~B. 1995, ApJ,
  454, 826

\reference Kramer, M. 1994, A\&AS, 107, 527

\reference Kramer, M., Wielebinski, R., Jessner, A., Gil, J.~A., \& Seiradakis, J.~H.
  1994, A\&AS, 107, 515



\reference Ray, P., van Kerkwijk, M.~H., Kulkarni, S.~R., Prince, T.~A., Sandhu, J.~S., \&
  Nice, D.~J. 1996, ApJ, 470, 1103

\reference Taylor, J.~H., \& Weisberg, J.~M. 1989, ApJ, 345, 434

\reference Taylor, J.~H., \& Cordes, J.~M. 1993, ApJ, 411, 674

\end{references}
\end{document}